\documentclass[11pt]{article}
\usepackage{amsmath}
\usepackage{amsgen,amstext,amsbsy,amsopn,amsthm, amssymb}

\newtheorem{thm}{Theorem}        
\numberwithin{thm}{section}
\newtheorem{cor}{Corollary}
\numberwithin{cor}{section}
\newtheorem{lem}{Lemma}
\numberwithin{lem}{section}

\numberwithin{prop}{section}
\theoremstyle{definition}

\newcommand{\beq}{\begin{equation}}
\newcommand{\eeq}{\end{equation}}
\newcommand{\beqa}{\begin{eqnarray}}
\newcommand{\eeqa}{\end{eqnarray}}

\newcommand{\var}{\varepsilon}

\newcommand{\vk}[1]{\mbox{{\bf #1}}}

\newcommand{\vks}[1]{\mbox{{\em {\bf #1}}}}

\newcommand{\uppi}[1]{\mbox{{\scriptsize #1}}}
\newcommand{\uppis}[1]{\mbox{{\em {\scriptsize #1}}}}

\def\A{{\bf A}}

\def\R{{\mathbb R}}
\def\r{{\bf x}}
\def\x{{\bf x}}
\newcommand{\X}{\bf X}
\def\C{{\mathbb C}}

\def\mfr#1/#2{\hbox{$\frac{{#1} }{ {#2}}$}}

\font\tenmb=cmmib10
\def\vsigma{\hbox{\tenmb\char27}}

\def\uprho{\raise1pt\hbox{$\rho$}}
\def\uprho{\raise1pt\hbox{$\rho$}}
\def\upchi{\raise1pt\hbox{$\chi$}}
\def\dlambda{\lower1pt\hbox{$\lambda$}}

\numberwithin{equation}{section}
\begin{document}

\title{\bf Asymptotic Exactness of Magnetic Thomas-Fermi Theory\\ at Nonzero Temperature}
\author{\vspace{5pt} Bergth\'or Hauksson$^{1*}$ and Jakob Yngvason$^{2}$\\
\small{$1.$ K\"ogun hf, 
Lyngh\'alsi 9, 
IS-110 Reykjavik, 
Iceland}\\ 
\small{$2.$ Institut f\"ur Theoretische Physik,
Universit\"at
Wien,}\\
\small{Boltzmanngasse 5, A-1090 Vienna, Austria}}
\date{\small 
$\phantom{x}$  \\
 {\it Dedicated to Elliott H.\ Lieb on the occasion of his 70th birthday}}
\maketitle
\renewcommand{\thefootnote}{$*$}
\footnotetext{Work partially
supported by the Research Fund of the University of Iceland.\\
\copyright\, 2003 by the authors. This paper may be reproduced, in its
entirety, for non-commercial purposes.}

\begin{abstract}
We consider the grand canonical pressure
for Coulombic matter with nuclear charges $\sim Z$ in a magnetic 
field $B$ and at nonzero temperature.
We prove that its asymptotic limit as $Z\to\infty$ with
$B/Z^3\to 0$ can be obtained by minimizing a Thomas-Fermi type pressure
functional.
\end{abstract}
\section{Introduction}

This paper intends to add one more chapter to the saga of rigorous
Thomas-Fermi theory in which Elliott H. Lieb played a prominent role
\cite{Sel}.  The issue is the derivation of Thomas Fermi theory at
nonzero temperature in a strong magnetic field from quantum
statistical mechanics.  The asymptotic exactness of Thomas-Fermi
theory for Coulombic matter in its ground state was first proved by
Lieb and Simon in the fundamental paper \cite{LiebSim}.  A shorter
proof, using coherent states, was given by Lieb in \cite{LiebTF}, and
several ideas in the present paper were inspired by that proof.

Thomas-Fermi Theory for matter in extremely strong magnetic fields is
important for the physics of neutron stars, cf.\ \cite{F1,F2, F3} and
references quoted therein.  This theory was analyzed from the point of
view of mathematical physics in \cite{JY,LSYb, I1} and its status as a
limit of quantum mechanics in a certain parameter range firmly
established; an extension of the asymptotics to inhomogeneous magnetic
fields is in \cite{ES}.  All these works are concerned with the ground
state, but non-magnetic TF theory at temperatures $T>0$ has been
treated in \cite{NT,T,M}.  Magnetic TF theory at nonzero temperature
was studied in \cite{R} and applied to the equation of state for
matter in the crust of a neutron star, but a proof of its asymptotic
exactness has, to the best of our knowledge, not been published until
now.  The proof we give here brings together techniques from
\cite{NT,T} and \cite{JY,LSYb} with several amendments and additions.

We start our discussion with some heuristic considerations.  A
possible point of departure for a motivation of TF theory, both at
$T=0$ and $T>0$, is the thermodynamic relation between the particle
density $\rho$, the chemical potential $\mu$, and the pressure $P$ for
a homogeneous gas of noninteracting particles (electrons):
\begin{equation}\label{murho} \rho = \partial P(\mu)/\partial\mu
    =: P'(\mu).  \end{equation}
The next step is to consider electrons that interact with with each
other via Coulomb forces and also with an external potential $V$
(arising from nuclei in fixed positions as well as a confining
potential that prevents the electrons from escaping to infinity).  The
electron density now depends on the position $\r$.  The TF theory is
formally obtained from (\ref{murho}) by replacing the constant density
$\rho$ by a position dependent density $\rho(\r)\geq 0$ and the
chemical potential by a position dependent chemical potential
$\mu(\r)$, imposing as an equilibrium condition that the {\it total}
electrochemical potential \begin{equation}\mu_{\rm
TF}=\mu(\r)+V_\rho(\r),\end{equation} with \begin{equation}
V_\rho(\r)=V(\x)+\rho *|\r|^{-1},\end{equation} should be independent
of $\r$.  The result is the Thomas-Fermi equation:
\begin{equation}\label{gentfeq}\rho(\r)=P'\left(\mu_{\rm
TF}-V_\rho(\r)\right).\end{equation} For given $P$, $V$ and $\mu_{\rm
TF}$ this is a nonlinear integral equation for $\rho(\r)$.  The total
particle number is \begin{equation}\label{subs}
\int\rho(\x)d\x=N.\end{equation}

The equation (\ref{gentfeq}) is the variational equation associated
with the minimization problem for the TF {\it pressure functional} of
the density,\footnote{Instead of considering $\mathcal P$ as a
functional of the density, it could equivalently be considered as a
functional of the potential $V_{\rho}$.  Note that $\rho$ and hence
$D(\rho,\rho)$ is determined by $V_{\rho}$ because $4\pi
\rho(\x)=-\Delta\rho*|\x|^{-1}=\Delta(V-V_{\rho}(\x))$.  While this
point of view (which is related to that of Firsov \cite{F} in standard
TF theory) may be more natural if $\mathcal P$ is regarded as a
Legendre transformation of the free energy functional $\mathcal F$
(\ref{gentffunc2}), we find it convenient in the present context to
regard ${\mathcal P}$ as a functional of $\rho$.  }
\begin{equation}\label{gentffunc}{\mathcal P}[\rho]=\int
P\left(\mu_{\rm TF}-V_{\rho(\r)}\right)d\x+D(\rho,\rho),\end{equation}
with \begin{equation} D(\rho,\rho)=
\frac12\int\int\frac{\rho(\r)\rho(\r')}{|\r-\r'|}d\r \,
d\r'.\end{equation} The minimum of (\ref{gentffunc}) over all
nonnegative functions $\rho$ will be called the TF {\it pressure}.

An alternative form of the TF equation is obtained if one replaces $P$
by its Legendre transform, the free energy density
\begin{equation}\label{legr}
f(\rho)=\sup_{\mu}\{\mu\rho-P(\mu)\}.\end{equation} From (\ref{legr})
it follows that $\partial f/\partial\rho=: f'$ is the inverse of $P'$. 
Hence the TF equation (\ref{gentfeq}) can also be written
\begin{equation}\label{gentfeq2}f'(\rho(\r))+V_{\rho}(\r))=\mu_{\rm
TF}.\end{equation} This form of the TF equation is associated with the
minimization problem for the {\it free energy functional}
\begin{equation}\label{gentffunc2}{\mathcal F}[\rho]=\int
\left\{f(\rho(\r))+V(\r))\rho(\r)\right\}d\x+D(\rho,\rho)\end{equation}
with (\ref{subs}) as a subsidiary condition and $\mu_{\rm TF}$ as a
Lagrange multiplier.

In the sequel we shall investigate the functional
(\ref{gentffunc}).  Our main result is its asymptotic exactness in the
case of electrons in a constant magnetic field and at nonzero
temperature.  This amounts to taking a semiclassical limit of the grand
canonical pressure for the quantum mechanical many-body problem.  The
corresponding problem for the free energy, i.e., the canonical
partition function, is technically more involved and will not be
tackled here.

We now introduce some notation that will allow us to state the results 
precisely.

The many-body Hamiltonian considered in this paper is
\begin{equation}\label{ham}H_{N,Z,B}=\sum_{i=1}^N\left\{ [{\bf
p}_{i}+\A(\r_i))\cdot{\vsigma}_i]^2
+V_{Z,B}({\r_{i}})\right\}+\sum_{1\leq i<j\leq N}\vert
\r_i-\r_j\vert^{-1}\end{equation} Here ${\bf p}=-{\rm i}\nabla$, ${\bf
A}(\r)=\mfr1/2 (-Bx_{2},Bx_{1},0)$ is the vector potential of a
homogeneous magnetic field of strength $B$ in the $x_{3}$-direction,
and ${\vsigma}$ is the vector of Pauli matrices.  Atomic units are
chosen so that $\hbar=2m=e=1$, and the temperature unit is such that
the Boltzmann constant $k$ is also 1.  The external potential is
\begin{equation} V_{Z,B}({\r}) = -Z\sum_{k=1}^K \frac{z_k}{|{\r} -
\ell{\X}_k|} +Z \ell^{-1}W(\ell^{-1}\r)\end{equation} where $W$ is a
confining potential that tends sufficiently rapidly to $\infty$ for
$|\x|\to\infty$ so that $\exp (-W(\x)/T)$ is integrable for all $T>0$. 
It will also be assumed to satisfy some regularity conditions stated
later.  The ${\X}_k$ are fixed positions of nuclei with fixed charges
$z_{k}\leq 1$ which are scaled by an overall parameter $Z$.  The
length scaling factor
\begin{equation}\label{ellbz}\ell=\ell_{Z,B}=Z^{-1/3}[1+(B/Z^{4/3})]^{-2/5}
\end{equation}
is the one appropriate for TF atoms in a magnetic field, cf. 
\cite{LSYb}.

The Hamiltonian $H_{N,Z,B}$ operates on the $N$-electron Hilbert space
of antisymmetric wave functions in space and spin variables:
\begin{equation} {\mathcal
H}_{N}=\wedge^{N}L^{2}(\R^{3},\C^{2}).\end{equation} The corresponding
Fock space is \begin{equation}{\widehat {\mathcal
H}}=\bigoplus_{N=0}^\infty {\mathcal H}_{N}\end{equation} with
${\mathcal H}_{0}=\mathbb C$.  If $A_{N}$ are operators on ${\mathcal
H}_{N}$, $N=0,1,\dots$, we denote the operator $\oplus_{N=0}^\infty
A_{N}$ on ${\widehat {\mathcal H}}$ by $\widehat A$.  In particular,
\begin{equation}\label{hzbhat}\widehat
H_{Z,B}=\bigoplus_{N=0}^{\infty} H_{N,Z,B}.\end{equation} The grand
canonical pressure at chemical potential $\mu$ and temperature $T$
is\footnote{Note that $\mu$ denotes here the total electrochemical
potential and not the chemical potential of the free electron gas
denoted by the same letter in in (\ref{murho}) and (\ref{ptb1}).}
\begin{equation} P^{\rm QM}(\mu,B,T,Z)=T\,\ln \hbox{\rm
tr}\exp[-(\widehat H_{Z,B}-\mu\widehat N)/T].\end{equation}

The free 1-particle Hamiltonian \begin{equation} H_{0}=[({\bf
p}+\A(\r))\cdot{\vsigma}]^2\end{equation} has the Landau spectrum
\begin{equation}\label{landauspec} \varepsilon_{\nu}(p)=2B\nu
+p^2\,,\quad\nu=0,1,\dots,\ p\in{\mathbb R}\end{equation} with degeneracy (pro
unit area in the $(x_{1},x_{2})$-plane)
\begin{eqnarray*}
  d_{\nu}(B) =  \left\{ \begin{array}{ccc}
             {B}/({2\pi}) & \mbox{{\rm if}} & \nu = 0 \\
           {B}/{\pi} & \mbox{{\rm if}} & \nu \geq 1.
             \end{array}
             \right.
\end{eqnarray*}
The pressure of a free electron gas at temperature $T$ in a magnetic
field and with chemical potential $\mu$ is
\begin{equation}\label{ptb1} P_{T,B}(\mu) = T\sum_{\nu =
0}^{\infty}d_{\nu}(B)\int_{-\infty}^{\infty}\ln\left[1 +
\exp\{-(\varepsilon_{\nu}(p) - \mu)/T \} \right]\, dp.\end{equation}
The {\it magnetic Thomas-Fermi pressure functional} is obtained by
taking $P$ in (\ref{gentffunc}) to be (\ref{ptb1}):
\begin{equation}\label{mtffunc}{\mathcal P}^{\rm
MTF}[\rho;Z,\mu,T,B]=\int P_{T,B}(\mu-V_{Z,B,\rho}({\r}))\;d^3{\r}
+D(\rho,\rho) \end{equation} with \begin{equation}
V_{Z,B,\rho}(\r)=V_{Z,B}(\r)+\rho *|\r|^{-1}.\end{equation} The
pressure according to MTF theory is \begin{equation} P^{\rm
MTF}(Z,\mu,T,B):=\inf_{\rho\geq 0} {\mathcal P}^{\rm
MTF}[\rho;\mu,T,B,Z].\end{equation}

We can now state our main result:

\begin{thm}[The MTF pressure is a limit of the QM pressure]
    If  $Z,\mu,T\to\infty$ with $\mu/(Z\ell_{B,Z}^{-1})$ and
$T/(Z\ell_{B,Z}^{-1})$ fixed but $B/Z^3\to 0$, where $\ell_{B,Z}$ is 
given by (\ref{ellbz}),
then
\begin{equation}\frac{P^{\rm QM}(\mu,T,B,Z)}{P^{\rm MTF}(\mu,T,B,Z)}\to 1
\end{equation}
    \end{thm}
The main steps in the proof are as follows.  In the next section we
discuss the MTF functional in more detail, in particular the existence
and uniqueness of a minimizer and its properties.  In Section 3 we
consider a Hamiltonian with a mean field and the corresponding
pressure functional.  Section 4 contains the basic semiclassical limit
theorem, which is proved using the magnetic coherent states introduced
in \cite{JY} and \cite{LSYb}.  As a corollary one obtains the
convergence of the mean field pressure to the MTF pressure.  Here it
is important that the error bounds in the semiclassical theorem are
uniform in the relevant parameters.  In the last section the proof of
Theorem 1.1 is completed by estimating the many-body pressure in terms
of the mean field pressure.
    
    \section{The MTF pressure functional}
    
	In this section we collect some basic facts about the pressure
	functional (\ref{mtffunc}).  We start with some formulas and
	estimates for the pressure (\ref{ptb1}) of the free electron
	gas in a magnetic field.  It can also be written as \begin{equation}
	P_{T,B}(\mu)\label{ptbg}= \int_{0}^\infty\frac
	{G(\varepsilon)}{\exp\{(\varepsilon-\mu)/T\}+1}d\varepsilon
	\end{equation} with the integrated density of states \begin{equation}
	G(\varepsilon)=\frac B{2^{1/2}\pi^2}\left[\varepsilon^{1/2}+2
	\sum_{\nu = 0}^{\infty}|\varepsilon-2B\nu|_{+}^{1/2}\right]. 
	\end{equation} Here $|s|_{+}:=|s|$ for $s\geq 0$ and 0 otherwise.  The
	corresponding formulas for $P'_{{T,B}}=\partial
	P_{{T,B}}/\partial\mu$ are \begin{eqnarray}\label{ptbder} P'_{T,B}(\mu)
	&= &\sum_{\nu =
	0}^{\infty}d_{\nu}(B)\int_{-\infty}^{\infty}\frac 1{
	\exp\{(\varepsilon_{\nu}(p) - \mu)/T \} +1}\, dp\\
	\label{ptbder2}&=& \int_{0}^\infty\frac
	{G'(\varepsilon)}{\exp\{(\varepsilon-\mu)/T\}+1}d\varepsilon. 
	\end{eqnarray} The scaling properties of $P_{{T,B}}$ can be seen by
	writing \begin{equation}\label{magnfreepress} P_{{T,B}}(\mu) = \frac{B
	T^{3/2}}{2^{1/2}\pi^2} \left[ I_{1/2} \left({\mu}/{ T} \right)
	+ 2 \sum_{\nu =1}^\infty I_{1/2} \left( {\mu - (2B\nu/T)}
	\right)\right], \end{equation} and \begin{equation}\label{magnfreedens}
	P'_{{T,B}}(\mu) = \frac{B T^{1/2}}{2^{3/2}\pi^2} \left[
	I_{-1/2} \left({\mu}/{ T} \right) + 2 \sum_{\nu =1}^\infty
	I_{-1/2} \left( {\mu - (2B\nu/T)} \right)\right] \end{equation} with
	\begin{equation} I_k(x) = \int_0^\infty \frac{y^k}{e^{y-x} + 1} dy.  
	\end{equation}
	Using the inequalities $\mfr1/2 \theta(-s)\leq
	(e^s+1)^{-1}\leq \theta(-s)+e^{-|s|})$ we obtain from
	(\ref{ptbg}) the following simple estimates, treating the sum
	over $\nu\geq 1$ as a Riemannian approximation for an
	integral:
	\begin{multline}
       c\left( B |\mu|_{+}^{3/2}+  
        |\mu|_{+}^{5/2} \right) \leq P_{{T,B}}(\mu) \\ 
	\leq C\left( B |\mu |_{+}^{3/2}+ 
        |\mu|_{+}^{5/2} + e^{-|\mu|/T} ( BT^{3/2}+ 
        T^{5/2}) \right)
        \label{ojafna9x}
\end{multline}
with constants $c>0$ and $C< \infty$.   In the same way we obtain
\begin{multline}
       c'\left(B |\mu|_{+}^{1/2}+ 
        |\mu |_{+}^{3/2}\right)\leq  P'_{{T,B}}(\mu)\\ 
	\leq C'\left(B |\mu|_{+}^{1/2}+ 
        |\mu |_{+}^{3/2} + e^{-|\mu|/T} ( B T^{1/2}+ 
        T^{3/2} )\right).
        \label{ojafna10x}\end{multline}

    As {\it domain of definition} for the magnetic pressure functional
    we take \begin{equation} {\mathcal M}=\{\rho:\, \rho(\x)\geq 0,\,
    D(\rho,\rho)<\infty\}.  \end{equation} To see that ${\mathcal P}^{\rm
    MTF}[\rho;\mu,T,B,Z]$ is well defined and $<\infty$ for $\rho\in
    {\mathcal M}$ we use (\ref{ojafna9x}) together with
    \begin{lem}[Coulomb bound]\label{L0}
If $\rho\geq 0$ with $D(\rho,\rho)<\infty$, then the potential
$v_{\rho}(\x)=\rho*|\x|^{-1}$ is in $L^6(\R^3)$; in fact
\begin{equation}\label{v6}\Vert v_{\rho}\Vert_{6}^2\leq {\rm 
(const.)} D(\rho,\rho).
\end{equation}
    \end{lem}
    \begin{proof}
Since $4\pi\rho=-\nabla^2 v_{\rho}$, we can write $D(\rho,\rho)={\rm 
(const.)}\Vert \nabla v_{\rho}\Vert_{2}^2$. We then use the Sobolev 
inequality in $\R^3$ \cite{LL}, i.e.,  $\Vert f\Vert_{6}\leq {\rm 
(const.)}\Vert \nabla f\Vert_{2}$.	
\end{proof}
From this lemma follows that $\mu-V_{Z,B,\rho}\in L^{5/2}(\R^3)_{\rm
loc}$ and hence also in $L^{3/2}(\R^3)_{\rm loc}$ for
$\rho\in{\mathcal M}$.  Moreover, since $W(\x)$ tends to $\infty$ if
$\x\to\infty$ while the negative potential from the nuclei tends to
zero, there is an $R\geq 0$ such that $|\mu-V_{Z,B,\rho}|_{+}=0$ and
$|\mu-V_{Z,B,\rho}|\geq Z\ell^{-1}W(\ell^{-1}\x)-{\rm (const.)}$ for
$|\x|\geq R$.  The finiteness of ${\mathcal P}^{\rm
MTF}[\rho;\mu,T,B,Z]$ now follows from (\ref{ojafna9x}) by splitting
the integration domain in (\ref{mtffunc}) into $|\x|\leq R$ and $|\x|>
R$.  \medskip

The MTF pressure functional is nonnegative and strictly convex since
$D(\rho,\rho)$ is strictly convex, $\rho\mapsto v_{\rho}$ is linear
and $s\mapsto \ln(1+\exp (-s))$ strictly convex.  Moreover, if
${\mathcal M}$ is equipped with the topology defined by the Hilbert
norm $D(\rho,\rho)^{1/2}$, then ${\mathcal P}^{\rm MTF}$ is weakly
lower semicontinuous.  This can be seen by representing the convex
functional as a supremum of affine, weakly continuous functionals in a
similar way as in \cite{T}, (4.1.10).  This continuity and strict
convexity implies by standard arguments (c.f., e.g. \cite{LiebTF})
that the functional has a {\it unique minimizer} $\rho^{\rm MTF}\in
{\mathcal M}$ (depending on the parameters $\mu,T,B,Z$).  It is the
unique solution to the {\it MTF equation}, i.e., the variational
equation for the minimization problem, \begin{equation}
\rho(\x)=P'_{T,B}(\mu-V_{Z,B,\rho}(\x)).\end{equation}

   Eq.\ (\ref{magnfreepress}) implies the following {\it scaling
   property} of the MTF functional: \begin{equation}\label{pmtfscaled} {\mathcal
   P}^{\rm MTF}[\rho;\mu,T,B,Z]=Z^{2}\ell^{-1} {\tilde{\mathcal
   P}}^{\rm MTF}[\tilde\rho;\tilde\mu,\tilde T,\beta]\end{equation} where 
   \begin{equation}
   \beta=B/Z^{4/3}\,,\quad\ell=Z^{-1/3}(1+\beta)^{-2/5},\end{equation} 
   \begin{equation}
   \label{rhoscaling}\rho(\x)=Z\ell^{-3}\tilde\rho(\ell^{-1}\x),
\end{equation}
   \begin{equation}\label{tildeT}\tilde \mu=\mu/(Z\ell^{-1})\,,\quad \tilde
   T=T/(Z\ell^{-1})\end{equation} and \begin{equation}\label{tildepfunc} 
   {\tilde{\mathcal
   P}}^{\rm MTF}[\tilde\rho;\tilde\mu,\tilde T,\beta]=
   (1+\beta)^{-3/5}\int P_{\tilde T,\tilde B}(\tilde \mu-\tilde
   V_{\tilde\rho}(\x))d\x+D(\tilde\rho,\tilde\rho)\end{equation} with 
   \begin{equation}\tilde
   B=B/(Z\ell^{-1})=\beta(1+\beta)^{-2/5},\end{equation} and 
   \begin{equation}\label{tildeV}
   \tilde V_{\tilde\rho}({\r}) = -\Phi_{\rm C}(\x)
   +W(\r)+\tilde\rho*|\x|^{-1}\end{equation} with \begin{equation} \Phi_{\rm
   C}(\x)=\sum_{k=1}^K \frac{z_k}{|{\r} - {\X}_k|}.\end{equation}

    We can also include the limiting cases $\beta=0$ and
    $\beta=\infty$: The case $\beta=0$ is just the temperature
    dependent TF without magnetic field considered in \cite{NT,T,M},
    while $\beta\to\infty$ means that only the lowest Landau level
    contributes.  It is the $T$ dependent version of the STF theory in
    \cite{LSYb}.  In fact, as is easily seen from
    (\ref{magnfreepress}), \begin{equation}\lim_{\beta\to\infty}(1+\beta)^{-3/5}
    P_{\tilde B,\tilde T}(\mu)=\frac{{\tilde
    T}^{3/2}}{2^{1/2}\pi^2}I_{1/2}(\mu/\tilde T)=: P^{\infty}_{\tilde
    T}(\mu)\end{equation} and hence \begin{equation}
    \label{betainf}\lim_{\beta\to\infty}{\tilde{\mathcal P}}^{\rm MTF}
    [\tilde\rho;\tilde\mu,\tilde T,\beta]=\int P^{\infty}_{\tilde
    T}(\tilde \mu-\tilde
    V_{\tilde\rho}(\x))d\x+D(\tilde\rho,\tilde\rho)=: {\tilde{\mathcal
    P}}^{\infty} [\tilde\rho;\tilde\mu,\tilde T]\end{equation}
    
    As last topic in this section we derive some uniform bounds for
    the minimizing density of the MTF functional and the corresponding
    Coulomb potential.  These bounds will be needed for the
    semiclassical limit theorem in Section 4.  \begin{lem}[Bounds for
    MTF minimizer]\label{L1} Let $\tilde\rho_{\beta}$ be the minimizer
    of $\tilde\rho\mapsto \tilde{\mathcal P}^{\rm
    MTF}[\tilde\rho;\tilde\mu,\tilde T,\beta] $ for fixed
    $\tilde\mu,\,\tilde T$.  Then
       
       \noindent
       (i) The potentials 
       $v_{\beta}(\x)=\tilde\rho_{\beta}*|\x|^{-1}$ are 
       bounded in $L^6(\R)$ uniformly in $\beta$.\\
       (ii) $\tilde\rho_{\beta}\in L^{p}(\R^3)$ for $1\leq p<2$ and
       $\Vert \tilde\rho_{\beta}\Vert_{p}$ is uniformly bounded in
       $\beta$.\\
       (iii) Let $j_{r}=r^{-3}j(\x/r)$ where $0\leq j\in C^\infty_{0}(\R^3)$ 
       satisfies $\int j(\x)d\x=1$. Then for all $R<\infty$ and $p<3$
       \begin{equation} \int_{|\x|\leq R} 
	   |v_{\beta}(\x)-v_{\beta}*|\x|^{-1}|^{p}d\x\to 0\end{equation}
       uniformly in $\beta$, as $r\to 0$.
      \end{lem}
      \begin{proof}
(i) This follows from Lemma \ref{L0} and the bound \begin{equation}
D(\tilde\rho_{\beta},\tilde\rho_{\beta})\leq {\tilde{\mathcal P}}^{\rm
MTF}[\tilde\rho_{\beta};\tilde\mu,\tilde T,\beta]\leq {\tilde{\mathcal
P}}^{\rm MTF}[0;\tilde\mu,\tilde T,\beta].  \end{equation} The right side is
continuous in $\beta\geq 0$ and converges to ${\tilde{\mathcal
P}}^{\infty} [0;\tilde\mu,\tilde T]<\infty$ for $\beta\to\infty$. 
Hence $D(\tilde\rho_{\beta},\tilde\rho_{\beta})$ is uniformly bounded
in $\beta$.

(ii) The MTF equation for the scaled density is \begin{equation}
\tilde\rho_{\beta}(\x)=(1+\beta)^{-3/5}P'_{\tilde T,\tilde B} (\tilde
\mu+\Phi_{C}(\x)-W(\x)-\tilde\rho_{\beta}*|\x|^{-1}).\end{equation} Now, since
$P'$ is monotonously increasing, the right side is bounded by
\begin{multline}(1+\beta)^{-3/5}P'_{\tilde T,\tilde B}
(\tilde \mu+\Phi_{C}(\x)-W(\x))\\ 
\leq C\left(|\tilde \mu+\Phi_{C}(\x)-W(\x))|_{+}^{3/2}+\exp(-|\tilde 
\mu+\Phi_{C}(\x)-W(\x))|/\tilde T)\right)
\end{multline}
which is in $L^p(\R^3)$ for all $1\leq p<2$.\end{proof}
          
(iii) This is proved in the same way as Proposition 4.19 (iii) in
\cite{LSYb}, using Jensen's and Young's inequalities together with (i)
and (ii).  Note that the Coulomb potential $|\x|^{-1}$ is in
$L^p(\R^3)_{{\rm loc}}$ for $p<3$.

   \section{Mean field theory}
  
   For $\rho\in {\mathcal M}$ we define a {\it mean field Hamiltonian}
   by \begin{equation} \label{mfham}H_{Z,B,\rho}=[({\bf
   p}+\A(\r))\cdot{\vsigma}]^2+V_{Z,B,\rho}(\x)\end{equation} and a {\it mean
   field pressure functional} by \begin{equation} \label{mffunc} 
   {\mathcal P}^{\rm
   mf}[\rho;\mu,T,B,Z]= T{\rm
   tr}\ln\left[1+\exp\{-(H_{Z,B,\rho}-\mu)/T\}\right]+D(\rho,\rho).
\end{equation}
   Note that the first term is equal to \begin{equation} T\,\ln \hbox{\rm
   tr}\exp[-(\widehat H_{Z,B,\rho}-\mu\widehat N)/T]\end{equation} where
   $\widehat H_{Z,B,\rho}$ is the second quantization of $
   H_{Z,B,\rho}$.

By exactly the same methods as in \cite{T}, (4.1.10)--(4.1.13), one
can show that (\ref{mffunc}) is strictly convex and weakly lower
semicontinuous on ${\mathcal M}$ and that the minimizer, $\rho^{\rm
mf}$, is the unique solution of the self-consistent (Hartree) equation
  \begin{equation}
        \rho(\vk{x}) = 2 \left\langle \vk{x}\left|
        \left( \exp\{(H_{Z,B,\rho}-\mu)/T\}+1
        \right)^{-1} \right| \vk{x}\right\rangle.
        \label{hartree}
\end{equation}
The right side is here the diagonal of the integral kernel of the
trace class opertor $\left(\exp\{(H_{Z,B,\rho}-\mu)/T\}+1
\right)^{-1}$.  Next we introduce the unitary operator
   \begin{equation}
        \left( U_{\ell} \psi \right)(\vk{x}) =
        \ell^{-3/2}\psi(\ell^{-1}\vk{x})
        \label{Ummyndun}
\end{equation}
for $\psi\in L^2(\R^3;\C^2)$ with $\ell$ given by (\ref{ellbz}).  It
transforms the Hamiltonian according to \begin{equation}\label{uscaling}
U_{\ell}^{-1}H_{Z,B,\rho}U_{\ell}=(Z\ell^{-1})\tilde
H_{h,b,\tilde\rho} \end{equation} with \begin{equation}
\tilde H_{h,b,\tilde\rho}= \left[ (h
\vk{p} + b \vk{a}(\vk{x}))\cdot \vsigma\right]^{2}+\tilde
V_{\tilde\rho}(\x).  \end{equation} Here $\vk{a}(\vk{x}) =
\mfr{1}/{2}(-x_{2},x_{1}, 0)$, $\tilde V_{\tilde\rho}$ is defined by
(\ref{tildeV}), and
\begin{eqnarray}
    \label{hal} h & = & \ell^{-1/2} Z^{-1/2} = Z^{-1/3}(1 +
    \beta)^{1/5}, \\
        b & = & B \ell^{3/2} Z^{-1/2} =  Z^{1/3}\beta(1 + \beta)^{-3/5}.
        \label{bal}
\end{eqnarray}
Since the trace is invariant under a unitary transformation we obtain
\begin{eqnarray}
    &T&\hskip-.2cm\mbox{tr}\,\ln(1 + \exp\{-(H_{Z,B,\rho}-\mu)/T\})
    =\\ \label{mfscaled} (Z\ell^{-1})\hskip-.2cm&\tilde
    T&\hskip-.2cm\mbox{tr}\,\ln(1 + \exp\{-( \tilde
    H_{h,b,\tilde\rho}-\tilde \mu)/\tilde T\}),
\end{eqnarray}
where $\tilde T$ and $\tilde\mu$ are given by (\ref{tildeT}).  Hence,
\begin{eqnarray} \label{mffunc2} \hskip-0,7cm{\mathcal P}^{\rm
mf}[\rho;\mu,T,B,Z]&\hskip-.2cm=\hskip-.2cm& Z\ell^{-1}\tilde
T\mbox{tr}\,\ln(1 + \exp\{-(\tilde H_{h,b,\tilde\rho}- \tilde
\mu)/\tilde T\})+Z^2\ell^{-1}D(\tilde\rho,\tilde\rho)\nonumber\\
&\hskip-.2cm=\hskip-.2cm&Z^2\ell^{-1}\left[Z^{-1}\tilde
T\mbox{tr}\,\ln(1 + \exp\{-(\tilde H_{h,b,\tilde\rho}- \tilde
\mu)/\tilde T\})+D(\tilde\rho,\tilde\rho)\right]\end{eqnarray} In the
next section we study the semiclassical limit $h\to 0$ of
(\ref{mffunc2}), which is equivalent to $Z\to\infty$ with $B/Z^3\to
0$.  The minimizer depends on $b$ and $h$, and in order to obtain a
semiclassical limit for the minima we need uniform bounds on the
minimizers and the corresponding Coulomb potentials.  The following
lemma is analogous to Lemma {\ref{L1}}.  \begin{lem}[Bounds for mean
field minimizer]{\label{L2}} Let $\tilde\rho_{b,h}$ be the minimizer,
for fixed $\tilde\mu,\,\tilde T$, of \begin{equation}\tilde{\mathcal P}^{\rm
mf}[\tilde \rho;\tilde \mu,\tilde T,b,h]=\left[Z^{-1}\tilde T\mbox{\rm
tr}\,\ln(1 + \exp\{-(\tilde H_{h,b,\tilde\rho}- \tilde \mu)/\tilde
T\})+D(\tilde\rho,\tilde\rho)\right].\end{equation} Then
    
   \noindent
       (i) The potentials 
       $v_{b,h}(\x)=\tilde\rho_{b,h}*|\x|^{-1}$ are 
       bounded in $L^6(\R)$ uniformly in $b,h$.\\
       (ii) $\tilde\rho_{b,h}\in  L^{1}(\R^3)$ and 
       $\Vert \tilde\rho_{b,h}\Vert_{1}$  is uniformly bounded 
       in $b,h$.\\
       (iii) Let $j_{r}=r^{-3}j(\x/r)$ where $0\leq j\in C^\infty_{0}(\R^3)$ 
       satisfies $\int j(\x)d\x=1$. Then for all $R<\infty$ and $p<3$
       \begin{equation} \int_{|\x|\leq R} 
	   |v_{b,h}(\x)-v_{b,h}*|\x|^{-1}|^{p}d\x\to 0\end{equation}
       uniformly in $b,h$, as $r\to 0$.
      \end{lem}
      \begin{proof}
(i) As in Lemma \ref{L1} the essential point is that 
$D(\tilde\rho_{b,h},\tilde\rho_{b,h})$ is uniformly bounded. In fact, 
\begin{equation} D(\tilde\rho_{b,h},\tilde\rho_{b,h})\leq
{\tilde{\mathcal P}}^{\rm mf}[\tilde\rho_{b,h};\tilde\mu,\tilde 
T,\beta]\leq {\tilde{\mathcal P}}^{\rm mf}[0;\tilde\mu,\tilde 
T,\beta].
\end{equation}
From the magnetic Lieb-Thirring inequality \cite{LSYb} it follows in 
the same way as in \cite{T}, (4.1,47), that the pressure
${\tilde{\mathcal P}}^{\rm mf}$
is, up to a constant factor,  bounded by the MTF pressure. The uniform 
bound thus follows in the same way as in Lemma \ref{L1} (i).\\
(ii) The Hartree equation for $\tilde \rho_{b,h}$ is
\begin{equation}
        \tilde\rho_{b,h}(\vk{x}) = 2 Z^{-1} \left\langle \vk{x}\left|
        \left( \exp\{ (\tilde H_{h,b,\tilde\rho_{b,h}}-\tilde 
        \mu)/\tilde T\}+1
        \right)^{-1} \right| \vk{x}\right\rangle.
        \label{hartreescal}
\end{equation}
Since $\tilde{\rho}*|\x|^{-1}\geq 0$, the integral of the right side
is bounded by the trace of the operator $2 Z^{-1} \left( \exp\{
(\tilde H_{h,b,0}-\tilde \mu)/\tilde T\}+ 1\right)^{-1}$, which can,
again by a magnetic Lieb-Thirring inequality, be bounded by the
corresponding semiclassical expression, i.e., $(1+\beta)^{-3/5}\int
(P'_{\tilde T,\tilde B} (\tilde \mu+\Phi_{C}(\x)-W(\x))d\x.$ This is
bounded uniformly in the parameters, by the same estimate as in Lemma
\ref{L1} (ii).\\
(iii) This is proved in the same way as in Proposition 4.19 in 
\cite{LSYb}, using (i) and (ii).
\end{proof}
    
\section{Semiclassics}

We consider generally the operator
\begin{equation}
        H_{h,b,v} = \left[(h\vk{p} +
        b\vk{a}(\vk{x}))\cdot{\vsigma}\right]^{2} + v(\vk{x})
\end{equation}
with $h>0$, $b \in \R$, $\vk{a}(\vk{x}) = \mfr{1}/{2}(-x_{2},
x_{1}, 0)$, and
\begin{equation}
        v(\vk{x}) = v_{1}(\vk{x}) + v_{2}(\vk{x})
\end{equation}
where $v_{1}\in L^{5/2}_{\rm loc}$ and $v_{2}$ is continuous with
$v_{2}(\vk{x}) \rightarrow \infty$ for $|\vk{x}| \rightarrow \infty$. 
We also impose some further conditions on $v_{1}$ and $v_{2}$ that are
described below.  In the application to (\ref{mfscaled}), \begin{equation}
v(\x)=\tilde V_{\tilde\rho}(\x)-\tilde\mu,\end{equation} and we shall take 
\begin{equation}
\label{v1}v_{1}(\x)=-\Phi_{\rm C}(\x)+\tilde\rho*|\x|^{-1}\end{equation} and
\begin{equation} v_{2}(\x)=W(\x)-\tilde \mu.  \end{equation}

The goal is to find an asymptotic approximation for
\begin{equation}
P^{\uppi{Q}}(h,b,v,\tau ):= \tau \mbox{tr} \ln \left( 1 +
\exp\left({-H^{1}_{h,b,v}/\tau}\right)\right)
        \label{qmthrystingur}
\end{equation}
as $h \rightarrow 0$.

The specific {\it conditions for $v_{1}$ and $v_{2}$} are:
\begin{enumerate}
    \item We assume that outside some compact ball,
    $B_{R_{0}}=\{\x:\,|\x|\leq R_{0}\}$, $v_{1}$ is subharmonic and
    the negative part, $|v_{1}(\x)|_{-}:=|v_{1}(\x)|$ for
    $v_{1}(\x)<0$ and 0 otherwise, is continuous and tends to zero at
    $\infty$.  This is fulfilled for (\ref{v1}) if
    $R_{0}>\max_{k}|\X_{k}|$.  Subharmonicity implies the following
    property that is convenient for the proof of the upper bound: Let
    $g\in C^{\infty}_{0}(\mathbb R^3)$ with $\int
    g^{2}(\vk{x})\,d\vk{x} = 1$ and define $g_{r}(\vk{x}) =
    r^{-3/2}g(\vk{x}/r)$ for $r>0$.  Then
    \begin{equation}\label{subharm}v_{1}(\x)-v_{1}*g_{r}^2(\x)\geq 0
    \end{equation} for
    $\x\notin B_{R_{0}}$ and $r$ small enough.
	
    \item  We assume that $v_{2}$ tends to
        $\infty$ sufficiently rapidly, so that $e^{-v_{2}(\cdot)/\tau} 
        \in L^{1}(\R^3)$ for all $\tau
        > 0$.  (This implies in particular that $e^{-v_{2}(\cdot)/\tau} 
	\in L^{p}(\R^3)$
       for all $p \geq 1$.)

        \item  We assume that $v_{2}$ is sufficiently regular so that 
        following holds: 
      Let $g\in C^{\infty}_{0}(\mathbb 
        R^3)$ with $\int g^{2}(\vk{x})\,d\vk{x} = 1$ and define 
	$g_{r}(\vk{x}) = r^{-3/2}g(\vk{x}/r)$ for $r>0$. Then we assume that 
	there exists a continuous function $v_{2}^{r}$ such that
        \begin{equation}\label{conv}
                v_{2}^{r} * g_{r}^{2} = v_{2}.
        \end{equation}
      Moreover, $\lim_{r\rightarrow 0}v_{2}^{r}(\vk{x}) = v_{2}(\vk{x})$
        for all $\vk{x}$, and $e^{-v_{2}^{r}(\cdot)/\tau} \leq f(\cdot,\tau)
        \in L^{1}(\R^3)$ for $r$ sufficiently small.  The same conditions 
        should be fulfilled for $v_{2,r} := v_{2} * g_{r}^{2}$.  
        
	These conditions on $v_{2}$ are not very restrictive. In fact,  
	Eq.\ (\ref{conv}) has a solution for all $C^{\infty}$ functions 
	$v_{2}$ by \cite{Hoerm}, Sect.\ 16.5, and by Fourier transform
	it easy to check 
	explicitly that the conditions hold, e.g., for all polynomials. 
	
\end{enumerate}

With $P_{T,B}(\mu)$ the pressure of the free electron gas, cf.\
(\ref{ptb1}) and (\ref{magnfreepress}), we define
\begin{eqnarray}\hskip-0.2cm
        P(w;h,b,\tau) &\hskip-0,2cm: =\hskip-0,2cm & h^{-3} P_{\tau, hb}(w)  \\
         &\hskip-0,2cm = \hskip-0,2cm& \tau\sum_{\nu =
        0}^{\infty}d_{\nu}(h,b)\int_{-\infty}^{\infty}\ln\left[1 +
\exp\{-((hp)^{2} +
        2 \nu hb - w)/\tau \} \right]\, dp
\end{eqnarray}
and
\begin{equation}
        P^{\uppi{scl}}(h,b,v,\tau) := \int P(-v(\vk{x});h,b,\tau)\,d\vk{x}.
\end{equation}

Then the following holds:
\begin{thm}[Semiclassical limit theorem]\label{sigild}
For fixed $\tau$ and $v$
\begin{equation}
        \lim_{h\rightarrow 0} \frac{P^{\uppis{Q}}(h,b,v,\tau
)}{P^{\uppis{scl}}(h,b,v,\tau)} = 1
        \label{sigiltmarkgildi}
\end{equation}
uniformly in $b$.
\end{thm}
\begin{proof}
We shall make use of convexity of the function
\begin{equation}
        \phi(s) = \tau\ln(1 + e^{-s/\tau}),
\end{equation}
which implies the inequality
\begin{equation}
        \phi(s+t) \geq \phi(s) + {\phi}'(s) t.
        \label{ojafna2}
\end{equation}
Moreover, if $L$, $M$ and $L+M$ are self-adjoint operators
then for all $\alpha\geq 1$
\begin{equation}
        \mbox{tr }\phi(L+M) \leq \mbox{tr }\phi(L) +
        \frac{1}{\alpha}\left[\mbox{tr } \phi(L+\alpha M) -\mbox{tr} \phi(L)
        \right].
        \label{trojafna}
\end{equation}

Another important tool for the proof are the magnetic coherent 
operators $\Pi(\nu,\vk{u},p)$ with $\nu = 0, 1, 2, \ldots$, 
$\vk{u} \in \R^{3}$
and $p \in \R$
introduced in \cite{JY} and \cite{LSYb}.  These operators fulfill the 
following conditions, cf.\  Eqs. (3.16)--(3.23) in  \cite{LSYb}:
\begin{eqnarray}
        \Pi(\nu,\vk{u},p) & \geq & 0. \\
        \sum_{\nu}\int \int\, \Pi(\nu,\vk{u},p)\,dp\, d\vk{u}& = & 1,
\label{haha}\\
        \mbox{tr}\,\Pi(\nu,\vk{u},p) & = & d_{\nu}(h,b)
        \label{pi4}, \\
        \mbox{tr}\, H_{h,b,v}\Pi(\nu,\vk{u},p)  & = &
        d_{\nu}(h,b)\left[\var_{p,\nu}(h,b) + v * g_{r}^{2}(\vk{u}) +
        \frac{h^{2}}{r^{2}} I_{g} \right],
        \label{jafna5}
\end{eqnarray}
with $\var_{p,\nu}(h,b) = (hp)^{2} + 2 b h \nu$ and $I_{g} =
\int(\nabla g)^{2}(\vk{x})\,d\vk{x}$.  Moreover,
\begin{eqnarray}\label{jafna418}
        H_{h,b,v} & = & \sum_{\nu}\int \int \left[\var_{p,\nu}(h,b) +
        v_{1}(\vk{u}) + v_{2}^{r}(\vk{u}) - \frac{h^{2}}{r^{2}} I_{g} \right]
        \Pi(\nu,\vk{u},p) \, d\vk{u} \, dp \nonumber \\
        & & \mbox{} + (v_{1} - v_{1} * g_{r}^{2} ).
        \label{jafna6}
\end{eqnarray}

We also use the following inequalities that are easy consequences of
(\ref{haha})--(\ref{pi4}) and convexity of $\phi$:
\begin{eqnarray}
        \mbox{tr}\, \phi(H_{h,b,v})  \geq  \sum_{\nu}\int \int
        d_{\nu}(h,b)\phi\left(\frac{1}{d_{\nu}(h,b)} \mbox{tr}
        \left(H_{h,b,v} \Pi(\nu,\vk{u},p) \right) \right)\, d\vk{u} \, dp  ;
        \label{ojafna7} \\
        \sum_{\nu}\int \int \phi\left(f(\nu,\vk{u},p)\right) \, d\vk{u} \, dp
        \geq  \mbox{tr}\,
         \phi\left(\sum_{\nu}\int \int f(\nu,\vk{u},p)\Pi(\nu,\vk{u},p)
         \, d\vk{u} \, dp\right),
        \label{ojafna8}
\end{eqnarray}
where Eq.\ (\ref{ojafna8}) holds for all sufficiently regular 
functions $f$ such that both sides are well defined.

The last preparatory step is to note that Eqs.\ (\ref{ojafna9x}) and 
(\ref{ojafna10x}) 
imply the following  estimates for
$P(w;h,p,\tau)$:
        \begin{multline}
	c\left( h^{-2} b |w|_{+}^{3/2}+ h^{-3} |w|_{+}^{5/2} \right)
	\leq P(w;h,p,\tau) \\ \leq C\left( h^{-2} b |w|_{+}^{3/2}+
	h^{-3} |w|_{+}^{5/2} + e^{-|w|/\tau} ( h^{-2} b \tau^{3/2}+
	h^{-3} \tau^{5/2}) \right)
        \label{ojafna9}
\end{multline}
and
\begin{equation}
        P'(w;h,p,\tau) \leq C'\left(h^{-2} b |w|_{+}^{1/2}+ h^{-3} 
        |w|_{+}^{3/2} + e^{-|w|/\tau} ( h^{-2} b \tau^{1/2}+ h^{-3}
        \tau^{3/2} )\right).
        \label{ojafna10}
\end{equation}
Eq.\ (\ref{ojafna9}) implies in particular that for fixed
$\tau$ and $v$
\begin{equation}
        P^{\uppi{scl}}(h,b,v,\tau) \sim (h^{-2}b + h^{-3}).
        \label{ojafna11}
\end{equation}

In order to prove (\ref{sigiltmarkgildi}) we thus have to estimate
$\tau \ln \left( 1
+ \exp(-H_{h,b,v}/\tau)\right)$ from above and below by
$P^{\uppi{scl}}(h,b,v,\tau)$ with errors that are small compared 
to (\ref{ojafna11}).

\subsection{Lower bound}
We use (\ref{jafna5}) and begin by writing
\begin{equation}
        \var_{p,\nu}(h,b)  +v * g_{r}^{2}(\vk{u})+ \frac{h^{2}}{r^{2}} I_{g}
        = A(\nu,\vk{u},p) + B(\nu,\vk{u},p)
\end{equation}
with
\begin{eqnarray*}
        A(\nu,\vk{u},p) & =  & \var_{p,\nu}(h,b) + v_{1}(\vk{u}) + v_{2} *
        g_{r}^{2}(\vk{u}),  \\
         B(\nu,\vk{u},p) & = &  \frac{h^{2}}{r^{2}} I_{g} +
         (v_{1}*g_{r}^{2}(\vk{u}) - v_{1}(\vk{u})).
\end{eqnarray*}
According to (\ref{ojafna7}) and (\ref{ojafna2}) we have
\begin{eqnarray*}
        P^{\uppi{Q}}(h,b,v,\tau) & \geq &  \sum_{\nu}\int \int
        d_{\nu}(h,b)\phi(A + B) \, d\vk{u} \, dp  \\
         & \geq & \sum_{\nu}\int \int
        d_{\nu}(h,b)\left\{ \phi(A) + {\phi}'(A) B \right\} \, d\vk{u} \, dp .
\end{eqnarray*}
We have to estimate the last term, i.e., the integral over ${\phi}'(A)
B$.  We use (\ref{ojafna10}) and the assumption that
$|v_{1}(\x)|_{-}\to 0$ and $v_{2}(\x)\to\infty$ if $|\x|\to\infty$, so
$|-v(\x)|_{+}=0$ for $\x$ outside some ball $B_{R}=\{\x:\, |\x|\leq
R\}$.  Since $v_{2}$ is continuous and $v_{2}*g_{r}^2$ therefore
bounded on $B_{R}$, uniformly in $r$, the terms involving $I_{g}$ are,
after division by $(h^{-2}b + h^{-3})$, bounded by
\begin{equation}\label{firstest} ({\rm const.})\left(\int_{|\x|\leq
R}|v_{1}(\x)|^{3/2}d\x+1+\int_{|x|>R}\exp(-v_{2}*g_{r}^2(\x)/\tau)d\x
\right) h^2/r^2.  \end{equation} Since $v_{1}\in L^{5/2}_{\rm loc}\subset
L^{3/2}_{\rm loc}$ and $\exp(-v_{2}*g_{r}^2/\tau)$ is bounded by an
$L^1$ function, independent of $r$, (\ref{firstest}) tends to 0 with
$h$ if $r=h^\delta$, $0<\delta<1$.  The terms involving
$v_{1}*g_{r}^{2}- v_{1}$ are, up to the factor $(h^{-2}b + h^{-3})$,
bounded by
\begin{multline}\label{secondest}
({\rm const.}) \int_{|\x|\leq
R}(|v_{1}(\x)|^{3/2}d\x+1)|\,|v_{1}*g_{r}^{2}(\x)- v_{1}(\x)|d\x \leq
\\ ({\rm const.}) \left( \int_{|\x|\leq R}(|v_{1}(\x)|^{5/2}d\x+1)|
d\x\right)^{3/5}\left(\int_{|\x|\leq R}|v_{1}*g_{r}^{2}(\x)-
v_{1}(\x)|^{5/2}d\x\right)^{2/5}
\end{multline}
which tends to zero with $r$ because $v_{1}\in L^{5/2}_{\rm loc}$.

\subsection{ Upper bound}
Here we use (\ref{jafna6}), (\ref{trojafna}) and (\ref{ojafna8}).  In 
addition we need the Lieb-Thirring inequality  for a constant 
magnetic field (see \cite{LSYb}), from which it follows in the same 
way as in
\cite{T}, Ex.\ 1 (4.1, 47), that
\begin{equation}
        P^{\uppi{Q}}(h,b,v,\tau) \leq (\mbox{const.}) 
	P^{\uppi{scl}}(h,b,v,\tau).
	\label{ojafna12}
\end{equation}
Now by (\ref{jafna6}) we can write $ H_{h,b,v}^{1} = L + M$ with
\begin{eqnarray*}
        L = \sum_{\nu}\int \int \left[\var_{p,\nu}(h,b) +
        v_{1}(\vk{u}) + v_{2}^{r}(\vk{u}) - \frac{h^{2}}{r^{2}} I_{g} \right]
        \Pi(\nu,\vk{u},p) \, d\vk{u} \, dp
\end{eqnarray*}
and
\begin{eqnarray*}
        M =  (v_{1} - v_{1} * g_{r}^{2} ).
\end{eqnarray*}
According to (\ref{trojafna}) we thus have
\begin{equation}
        P^{\uppi{Q}}(h,b,v,\tau) \leq \mbox{tr}\,\phi(L) + \frac{1}{\alpha}
        \left[\mbox{tr } \phi(L+\alpha M) -\mbox{tr} \phi(L)
        \right]
\end{equation}
for all $\alpha \geq 1$, and by (\ref{ojafna8}) we have
\begin{equation}
        \mbox{tr}\,\phi(L) \leq  P^{\uppi{scl}}(h,b,v_{1} + v_{2}^{r} -
        h^{2}I_{g}/r^{2},\tau).
        \label{ojafna13}
\end{equation}
If $r = h^{\delta}$, $0<\delta<1$, then it follows from the properties
of $v_{2}$ and the dominated convergence theorem that the right side
of (\ref{ojafna13}) converges to $ P^{\uppi{scl}}(h,b,v,\tau)$ (in the
sense that the ratio tends to 1), if $h \rightarrow 0$.

We thus have to show that it is possible to let  $\alpha \rightarrow \infty$
as $r \rightarrow 0$, in such a way that $\mbox{tr}\, \phi(L + \alpha M)$ 
stays bounded by $(\mbox{const.}) \cdot (h^{-3} + bh^{-2})$.

Now $L+\alpha M=H^1_{h,b,v}+(\alpha-1)L$ and thus \begin{equation}\mbox{tr}\,
\phi(L + \alpha M) = P^{\uppi{Q}}(h,b,v + (\alpha - 1)[v_{1} - v_{1} *
g_{2}^{r}], \tau).\end{equation} To estimate this we use the inequality
(\ref{ojafna12}) which gives
\begin{equation}
        \mbox{tr}\, \phi(L + \alpha M) \leq (\mbox{const.}) \int
        P^{\uppi{scl}}(h,b,-v(\vk{x}) - (\alpha - 1)[v_{1}(\vk{x}) - v_{1} *
        g_{r}^{2}(\vk{x}) ], \tau)\, d\vk{x}.
\end{equation}
We estimate this further using (\ref{ojafna9}) as well as 
the assumptions on $v_{1}$ and $v_{2}$, which 
imply that, outside some compact ball $B_{R}$,
\begin{equation} |-v(\vk{x}) - (\alpha - 1)[v_{1}(\vk{x}) - v_{1} *
        g_{r}^{2}(\vk{x})|_{+}=0\end{equation} and \begin{equation} |
	-v(\vk{x}) - (\alpha - 1)[v_{1}(\vk{x}) - v_{1} *
        g_{r}^{2}(\vk{x})|\geq |v_{2}(x)|-\mbox{(const.)}.\end{equation}
Now
\begin{equation}
\int_{|\x|\leq R}|v_{1}(\x) -  v_{1} * g_{r}^{2}(\x)|^{5/2}dx\leq 
\varphi(r)
\end{equation}
for some function $\varphi$ with $\varphi(r) \rightarrow 0$ if $r\rightarrow 0$.
 If we choose $\alpha = 1 + \varphi(r)^{-1}$ it follows from 
 (\ref{ojafna9}) that
$\mbox{tr}\,  \varphi(L + \alpha M) $ is bounded by $(\mbox{const.})\cdot
(h^{-3}  + bh^{-2})$ for all sufficiently small $r$.
\end{proof}

In the course of the proof we have shown that (for fixed $\tau$ and
$v$) \begin{equation} \left|P^{\rm Q}(h,b,v,\tau)-P^{\rm
scl}(h,b,v,\tau)\right|=o(h^{-3}+h^{-2}b)\end{equation} and hence, for $h$ and
$b$ defined in (\ref{hal}), (\ref{bal}), \begin{equation} \left|h^3P^{\rm
Q}(h,b,v,\tau)-\int P_{\tau, hb}(-v(\x)d\x\right|=
o(1+hb)=o((1+\beta)^{3/5}).\end{equation} Note also that $h^3=
Z^{-1}(1+\beta)^{3/5}$, $hb=\beta(1+\beta)^{-2/5}=\tilde B$.  We now
apply this to compare the MFT functional (\ref{mtffunc}) with the mean
field functional (\ref{mffunc}), making use of the scalings
(\ref{pmtfscaled}) and (\ref{mffunc2}).  At this point we need to
assume that the confining potential $W$ satisfies the regularity
conditions stated for $v_{2}$ above.  Without further ado we obtain
\begin{cor}\label{cor1}[Convergence of functionals]
   If $Z\to \infty$ while $B/Z^3\to 0$ and $\tilde\rho$, $\tilde 
   \mu$ and $\tilde T$ are fixed, then
   \begin{equation}\label{430}
   \left|Z^{-2}\ell\,{\mathcal P}^{\rm mf}[\rho;\mu,T,B,Z]-
   \tilde{\mathcal P}^{\rm MTF}[\tilde\rho;\tilde \mu,\tilde 
   T,\beta]\right|\to 0.\end{equation}
   \end{cor}
 
  For the application to Theorem 1.1, however, we need more than
  convergence of the functionals for fixed $\tilde \rho$, namely the
  convergence of the minima:\footnote{Pointwise convergence of
  functionals does in general not imply convergence of their infima,
  even if the functionals are strictly convex and their minimizers lie
  in a compact set.  This can be seen from the following example :
  Take ${\mathcal F}_{n}(x)=x_{n}+\Vert x\Vert^2$ on the Hilbert space
  $\ell^2$ of square summable sequences $x=(x_{1},x_{2},\dots)$.  Then
  ${\mathcal F}_{n}(x)\to \Vert x\Vert^2=:{\mathcal F}(x)$ for all $x$
  and the infimum of ${\mathcal F}$ is zero.  The infimum of
  ${\mathcal F}_{n}$, on the other hand, is $-1/4$.  All infima are
  attained in the weakly compact unit ball in $\ell^2$.}
    
\begin{cor}\label{cor2}[Convergence of minima]
   If $Z\to \infty$ while $B/Z^3\to 0$  and $\tilde 
   \mu$, $\tilde T$  are fixed, then
   \begin{equation}
   \left|Z^{-2}\ell\,{P}^{\rm mf}(\mu,T,B,Z)-
   \tilde{P}^{\rm MTF}(\tilde \mu,\tilde 
   T,\beta)\right|\to 0.\end{equation}
   \end{cor}
  \begin{proof}
       Let $\tilde\rho_{\beta}$ be the minimizer of $\tilde\rho\mapsto
       \tilde{\mathcal P}^{\rm MTF}[\tilde\rho;\tilde\mu,\tilde
       T,\beta] $ for fixed $\tilde\mu,\,\tilde T$.  Let
       $\rho_{\beta}$ be the corresponding unscaled density given by
       (\ref{rhoscaling}).  If $Z\to\infty$ with $\beta$ fixed (which
       implies $B/Z^3\to 0$) it follows from Corollary \ref{cor1} that
       \begin{eqnarray} Z^{-2}\ell{P}^{\rm mf}(\mu,T,B,Z)&\leq &Z^{-2}\ell\,
       {\mathcal P}^{\rm mf}[\rho_{\beta};\mu,T,B,Z]\nonumber\\ &\to &
       {\mathcal P}^{\rm MTF}[\tilde\rho_{\beta};\tilde \mu,\tilde
       T,\beta]=\tilde{P}^{\rm MTF}(\tilde \mu,\tilde T,\beta).\end{eqnarray}
       The general condition $B/Z^3\to 0$, however, allows $\beta$ and
       hence $\tilde\rho_{\beta}$ to vary as $Z\to \infty$ ($h\to 0$)
       so one must check that the error terms in the semiclassical
       proof are uniform in $\beta$.  These terms involve
       $\int_{|\x|\leq R}|v_{1}(x)-v_{1}*g_{r}^2(\x)|^{5/2}d\x$ with
       $v_1=-\Phi_{\rm C}+\tilde\rho_{\beta}*|\cdot|^{-1}$ and the
       required uniformity follows from Lemma \ref{L1}.  The converse
       inequality, \begin{equation} \lim Z^{-2}\ell\,{P}^{\rm mf}(\mu,T,B,Z)\geq
       \tilde{P}^{\rm MTF}(\tilde \mu,\tilde T,\beta), \end{equation} follows in
       the same way by noting that the minimizer $\tilde\rho_{h,b}$ of
       the mean field functional enters in in the error terms of the
       upper bound only through $L^p$-norms of
       $v_{h,b}=\tilde\rho_{h,b}*|\cdot|^{-1}$ and
       $v_{h,b}-v_{h,b}*g_{r}^2$ that are uniformly bounded by Lemma
       \ref{L2}.\end{proof}

\section{Proof of the QM limit theorem} 

In this section we complete the proof of Theorem 1.1. We use the 
notation explained in the Introduction, in particular 
$\widehat H_{Z,B}$ for the Hamiltonian (\ref{hzbhat}) on Fock space and
$\widehat H_{Z,B,\rho}$ for the second quantization of the mean field 
 Hamiltonian (\ref{mfham}).
\smallskip

\subsection{Upper bound}
We use the inequality \cite{LT}
\begin{equation}
        \sum_{i<j}^{N}\frac{1}{|\vks{x}_{i}-\vks{x}_{j}|} \geq
\sum_{i=1}^{N}\int_{{\mathbb R}^{3}}\frac{\rho(\vks{x})}{|\vks{x}-\vks{x}_{i}|}d
\vks{x}
        - D(\rho,\rho) - 3,68\gamma N -
        \frac{3}{5\gamma}\int_{{\mathbb R}^{3}}\rho^{5/3}(\vks{x}) \, d\vks{x}
	\label{TFnedramat}
\end{equation}
that holds for all $\gamma>0$ and all 
$\rho\in {\mathcal M}\cap L^1(\R^3)\cap L^{5/3}(\R^3)$.
Eq.\ (\ref{TFnedramat}) implies 
\begin{equation} \widehat H_{Z,B}\geq 
\widehat H_{Z,B,\rho}-3.68\gamma \widehat N-D(\rho,\rho)-
C_{\gamma,\rho}\end{equation}
with $C_{\gamma,\rho}=\frac{3}{5\gamma}\int_{{\mathbb R}^{3}}
\rho^{5/3}(\vks{x}) \, d\vks{x}$.
Since 
 $A \geq B$  implies $\mbox{tr }e^{-A} \leq \mbox{tr }e^{-B}$,
we get an upper bound on the grand canonical pressure:
\begin{eqnarray}
        P^{\uppi{QM}}(\mu, T, B, Z) & \leq & T \ln \left( \mbox{tr }
        e^{-(\widehat H_{Z,B,\rho}-\mu_{\gamma}\widehat N)/T} \right) +
         D(\rho,\rho) + C_{\gamma,\rho} \nonumber
 \\
         & = & T \mbox{tr }\ln(1 + e^{-(H_{Z,B,\rho}-\mu_{\gamma})/T})  +
         D(\rho,\rho) + C_{\gamma,\rho}\\
	 &=& {\mathcal P}^{\rm mf}
	 [\rho;\mu_{\gamma},T,B,Z]+C_{\gamma,\rho}
        \label{efraPmat}
\end{eqnarray}
where $\mu_{\gamma}:=\mu+3.68\gamma$.  We now apply the unitary
scaling (\ref{uscaling}) explained in Section 3 and note that \begin{equation}
\tilde{\mu_{\gamma}}=\tilde\mu+(Z^{-1}\ell) 3.68\gamma\end{equation} and 
\begin{equation}
C_{\gamma,\rho}=(Z^2\ell^{-1}) \frac{3}{5\gamma
Z^{1/3}\ell}\int\tilde\rho^{5/3}(\x)d\x.\end{equation} We choose $\gamma$ as a
function of $Z$ and $B$ such that $Z^{-1}\ell\gamma\to 0$ but
$Z^{1/3}\ell\gamma\to \infty$ if $Z\to \infty$, which is fulfilled as
long as \begin{equation} [1+(B/Z^{4/3})]^{2/5}\ll \gamma \ll Z^{4/3}
[1+(B/Z^{4/3})]^{2/5}.\end{equation} Then, according to \eqref{mffunc2} and
Corollary \ref{cor1}, we get as $Z\to \infty$ with $B/Z^3\to 0$ and
$\tilde\mu$, $\tilde T$ fixed: \begin{equation}\lim (Z^{-2}\ell)
P^{\uppi{QM}}(\mu, T, B, Z)\leq \tilde{\mathcal P} ^{\rm
MTF}[\tilde\rho;\tilde \mu,\tilde T,\beta]\end{equation} for each
$\tilde\rho\in{\mathcal M}\cap L^1(\R^3)\cap L^{5/3}(\R^3)$, in
particular for the minimizer $\tilde\rho_{\beta}$.  Note that
$\beta=\infty$ is allowed, c.f. Eq.\ \eqref{betainf}, and we can also
let $\tilde\rho=\tilde\rho_{\beta}$ vary for $\beta\to\infty$, because
$\Vert\tilde\rho_{\beta}\Vert_{5/3}$ is uniformly bounded by Lemma
\ref{L1} (ii).  Altogether, by \eqref{pmtfscaled}, \begin{equation}\lim
\frac{P^{\rm QM}(\mu,T,B,Z)}{P^{\rm MTF}(\mu,T,B,Z)}\leq 1\end{equation} as
$Z\to\infty$, uniformly in $B$ as long as $B/Z^3\to 0$.
\subsection{ Lower bound} 

We use the Peierls-Bogoliubov inequality (\cite{T}, (2.1.7), (2.1.8)): If 
$A$, $B$ and $A+B$ are self-adjoint operators
and $F(A) =
\ln \mbox{tr\,}e^{-A/T}$, then
\begin{equation}
	F(A+B) \geq F(A) - \langle B \rangle_{A}
\end{equation}
where
\begin{equation}
\langle B \rangle_{A} := \frac{\mbox{tr\,}\left(
Be^{-A/T}\right)}{\mbox{tr\, }e^{-A/T}}.
\end{equation}
We use this inequality with $A+B=\widehat{H}_{Z,B} - \mu \widehat{N}$
and $A=\widehat{H}_{Z,B,\rho}-\mu\widehat N-D(\rho,\rho)$. 
Then $B$ is, apart from the constant $D(\rho,\rho)$,
the second quantization of 
\begin{equation}
\sum_{i<j}^{N}|\vk{x}_{i} - \vk{x}_{j}|^{-1} -
        \sum_{i=1}^{N}\rho*|\vk{x}_{i}|^{-1}. 
	\end{equation}
In terms of the creation and annihilation operators 
$a_{s}^{*}(\vk{x})$ and $a_{s}(\vk{x})$ ($s$= spin component) this can be 
written
	\begin{eqnarray}
	B=\frac{1}{2}\sum_{s,s'}\int\int\,a_{s}^{*}(\vk{x})
        a_{s'}^{*}(\vk{x}')|\vk{x} - \vk{x}'|^{-1} a_{s'}(\vk{x}')
        a_{s}(\vk{x})\,d\vk{x}d\vk{x}'   \nonumber \\
       \mbox{}- \int \, \rho(\vk{y})
        \sum_{s,s'}\int a_{s}^{*}(\vk{x})a_{s}(\vk{x})
        |\vk{x} - \vk{y}|^{-1}\,d\vk{x}d\vk{y}+D(\rho,\rho)
	\end{eqnarray}

We thus get
\begin{equation}
        P^{\uppi{QM}}(\mu,B,Z,T) \geq T \ln \left( \mbox{tr\,}
        e^{-(\widehat{H}_{Z,B,\rho}-\mu\widehat N)/T} \right) 
        +D(\rho,\rho)- 
	T \langle B \rangle_{A}
\end{equation}
We want to show that we can choose $\rho$ such that 
$\langle B \rangle_{A}\leq 0$.

Since $A$ is the second quantization of a one-particle operator, all
expectation values of products of creation and annihilation operators
in the state $\langle \cdot \rangle_{A}$ can be written in terms of
two point correlations by using Wick's theorem.  The expectation value
$\langle B \rangle_{A}$ involves terms of the form
\begin{equation}
        \langle a_{s}^{*}(\vk{x})a_{s'}^{*}(\vk{x}')
           a_{s'}(\vk{x}')a_{s}(\vk{x}) \rangle_{A},
        \label{threnn}
\end{equation}
which by   can be written
\begin{eqnarray}
        \langle a_{s}^{*}(\vk{x})a_{s}(\vk{x}) \rangle_A
        \langle a_{s'}^{*}(\vk{x}')a_{s'}(\vk{x}') \rangle_A
-       \langle a_{s}^{*}(\vk{x})a_{s'}(\vk{x}') \rangle_A
        \langle a_{s'}^{*}(\vk{x}')a_{s}(\vk{x}) \rangle_A \nonumber \\
= \langle a_{s}^{*}(\vk{x})a_{s}(\vk{x}) \rangle_A
        \langle a_{s'}^{*}(\vk{x}')a_{s'}(\vk{x}') \rangle_A
 -      |\langle a_{s'}^{*}(\vk{x}')a_{s}(\vk{x}) \rangle_A|^{2}.
        \label{fern}
\end{eqnarray}
If we define
\begin{equation}
        \overline{\rho}(\x) = \sum_{s} \langle a_{s}^{*}(\vk{x})a_{s}(\vk{x})
\rangle_A
\end{equation}
then, using (\ref{fern}), we can write
\begin{eqnarray}
        \langle B \rangle_{A} & = &
D(\overline{\rho},\overline{\rho})
        - 2 D(\rho,\overline{\rho}) + D(\rho,\rho) -
        \frac{1}{2}\sum_{s,s'}\int\int\,
        \frac{|\langle a_{s'}^{*}(\vk{x}')a_{s}(\vk{x})
\rangle_A|^{2}}{|\vk{x} -
        \vk{x}'|}  \,d\vk{x}\,d\vk{x}'
        \nonumber \\
         & \leq & D(\rho-\overline{\rho},\rho-\overline{\rho}).
        \label{fim}
\end{eqnarray}
We now choose $\rho = \overline{\rho}$, i.e., we choose $\rho = 
\rho^{\rm mf}$
where $\rho^{\rm mf}$ is the solution of the Hartree equation
\eqref{hartree}.
This gives
\begin{equation}
        P^{\uppi{Q}}(\mu, Z, B, V) \geq
        P^{\uppi{mf}}(\mu; Z, B, T).
\end{equation}
By Corollary \ref{cor2}, $P^{\uppi{mf}}(\mu; Z, B, 
T)/P^{\uppi{MTF}}(\mu; Z, B, T)$ converges to 1 in the limit 
considered. Thus 
\begin{equation}\lim \frac{P^{\rm QM}(\mu,T,B,Z)}{P^{\rm MTF}(\mu,T,B,Z)}
    \geq 1\end{equation}
and the proof of Theorem 1.1 is complete.

\section{Conclusions} Using magnetic coherent states we have proved a
semiclassical limit theorem for a mean field quantum mechanical
pressure functional and applied it to derive MFT theory at nonzero
temperatures as a limit of quantum statistical mechanics in a certain
parameter range.  Our result concerns the grand canonical partition
function while the corresponding result for the free energy is left as
an open problem.  Other interesting questions not tackled here concern
the extension of the asymptotic ground state classification
\cite{LSYa} to temperature and field strength regions beyond those of
the present analysis, as well as the thermodynamic limit, where the
number of nuclei tends to infinity.  \medskip

\noindent {\bf Acknowledgements.} We thank Elliott Lieb, Heide
Narnhofer, Walter Thirring and Ragnar Sigurdsson for helpful remarks. 
J.Y. acknowledges the hospitality at the Science Institute of the
University of Iceland in Reykjavik where most of this work was carried
out.

\end{document}